\documentclass[aps,prl,msmath,amssymb,reprint]{revtex4-2}
\usepackage{graphicx}
\usepackage{hyperref}
\hypersetup{colorlinks=true}
\usepackage{siunitx}
\usepackage{pdfpages}
\usepackage{etoolbox} 

\usepackage{xcolor}
\definecolor{ddgreen}{RGB}{0,150,50}
\definecolor{ddcyan}{RGB}{0,150,180}
\hypersetup{colorlinks=true,
    linkcolor=blue,
    filecolor=magenta,      
    urlcolor=ddcyan,
    citecolor=ddgreen}

\def\ket#1{| #1 \rangle}

\makeatletter
\patchcmd{\@outputpage@head}{\@ifx{\LS@rot\@undefined}{}{\LS@rot}}{}{}{}
\makeatother

\begin{document}
\title{Autodetachment of diatomic carbon anions from long-lived high-rotation quartet states}
\author{Viviane~C.~Schmidt}
\affiliation{Max-Planck-Institut f\"{u}r Kernphysik, Saupfercheckweg 1,
69117 Heidelberg, Germany}
\author{Roman~\v{C}ur\'{\i}k} \email{roman.curik@jh-inst.cas.cz}
\affiliation{J. Heyrovsk\'{y} Institute of Physical Chemistry, ASCR,
Dolej\v{s}kova 3, 18223 Prague, Czech Republic}
\author{Milan On\v{c}\'{a}k}
\affiliation{Institut f\"ur Ionenphysik und Angewandte Physik, Leopold-Franzens-Universit\"{a}t 
Innsbruck, Technikerstraße 25/3, Innsbruck 6020, Austria}
\author{Klaus~Blaum}
\author{Sebastian~George}
\author{J\"urgen~G\"ock}
\author{Manfred~Grieser}
\author{Florian~Grussie}
\author{Robert~von~Hahn}
  \altaffiliation[]{Deceased}
\author{Claude~Krantz}
\author{Holger~Kreckel}
\author{Old\v{r}ich~Novotn\'{y}}
\author{Kaija~Spruck}
\author{Andreas~Wolf}
\affiliation{Max-Planck-Institut f\"{u}r Kernphysik, $\,$Saupfercheckweg 1,
69117 Heidelberg, Germany}
\date{\today}

\begin{abstract}
We show that strong molecular rotation drastically modifies the autodetachment of C$_2{}^{-}$ ions in the lowest quartet electronic state a$^4\Sigma^+_u$. 
In the strong-rotation regime, levels of this state only decay by a process termed “rotationally assisted” autodetachment, whose theoretical description is worked out based on the non-local resonance model. 
For autodetachment linked with the exchange of six rotational quanta, the results reproduce a prominent, hitherto unexplained electron emission signal with a mean decay time near 3 milliseconds, observed on stored  C$_2{}^{-}$ ions from a hot ion source.
\end{abstract}

\maketitle

Properties and decay mechanisms of highly excited small molecules attract wide interest in fields such as molecular astrophysics \cite{tielens_molecular_2013,chang_hydroxyl_2019}, atmospheric science \cite{chang_vacuum_2023}, optical control of chemical reactions \cite{venkataramanababu_enhancing_2023} and plasma processing \cite{oehrlein_foundations_2018}.
Thus, recently, extreme rotational excitation created by optical excitation was found to enhance the reactivity of trapped diatomic cations \cite{venkataramanababu_enhancing_2023}, while signatures of strongly rotating diatomic radicals were even found in astrophysical observations \cite{tappe_discovery_2008,chang_hydroxyl_2019}.
Diatomic anions with similarly high excitation are being studied using stored ion beams \cite{von_Hahn_RevSciInstr_2016,schmidt_first_2013,schmidt_negative_2020}.
Here, highly excited molecules are created by fast-ion sputtering from solid surfaces \cite{urbassek_sputtering_1986,anders_molecule_2015}, where they acquire rotational and vibrational energy up to the dissociation limit \cite{anders_atom_2013}.
Studies at ion storage rings \cite{menk_vibrational_2014,Fedor_PRL_2005,Anderson_PRL_2020,iida_state-selective_2020,*iida_correction_2021} revealed the stability properties of such anionic molecules when emitting either the excess electron (autodetachment) or a heavy constituent (autofragmentation).
Recently observed autodetachment on these systems \cite{Anderson_PRL_2020} was attributed to high vibrational excitation \cite{Anderson_PRL_2020,Jasik_Franz_AD_Ag_2021}.

The present work focuses on decay processes at high internal excitation for stored diatomic carbon anions C$_2{}^-$. 
Unique among the small molecular anions, C$_2{}^-$ has low-lying electronic states sharing with the ground state the doublet spin symmetry and enabling optical spectroscopy \cite{Rehfuss_JCP_1988} and even laser cooling \cite{lasercooling_c2-_yzombard}. 
Correspondingly, most of its higher vibrational levels decay by electronic radiative transitions, while rotational excitation is much more stable, a permanent dipole moment still being absent in the homonuclear system.
Hence, by storing the C$_2{}^-$ anions extracted from a hot source, properties of this molecule can be investigated at high, mainly rotational excitation.

In this Letter, we analyze the autodetachment of C$_2{}^-$ under such excitation conditions. 
Based on previous observations \cite{Anderson_Z_Phys_D_1997,Pedersen_JCP_1998,Iizawa_JPS_Japan_2022,unimol_remark} autodetachment of C$_2{}^-$ from a hot ion source leads to a prominent near-exponential decay signal with a time constant close to $\SI{3}{\milli\second}$, which could not be linked to rates of any competing radiative decay \cite{Pedersen_JCP_1998,Iizawa_JPS_Japan_2022}.
No conclusive explanation could so far be given for the process behind these observations.

We show that this temporal behavior likely indicates a drastic modification of autodetachment in the lowest quartet state of C$_2{}^-$ by strong molecular rotation.
At low rotation, the electronic energy of this state lies $\sim${\,}$\SI{4}{\electronvolt}$ above the anionic doublet ground state and $\sim${\,}$\SI{1.4}{\electronvolt}$ above the detachment energy to neutral C$_2$.
Direct diabatic autodetachment (AD) then leads to pico- to femtosecond lifetimes of any quartet levels.
However, rotation up-shifts the relevant minima of the electronic potentials more strongly for the neutral than for the anionic quartet states.
At sufficient rotation, quartet levels appear for which AD is energetically forbidden {\em at fixed rotation} and can only proceed when the rotational state changes during the AD process.
We develop the theoretical description of such processes, which we term “rotationally assisted” AD.
Our calculated rates show that AD from rotationally excited C$_2{}^-$ ions in the lowest quartet state, connected with a change of internal rotation by six quanta, well explains the observed AD signal.
The results also show that this AD signal represents a limited range of rotational excitation and that more stable quartet C$_2{}^-$ ions should exist at even stronger rotation.

The AD of vibrationally excited levels in the  A$^2\Pi_u$ and B$^2\Sigma_u^+$ states of C$_2{}^-$ was observed by electron spectroscopy \cite{Mead_C2_JCP_1985,Hefter_C2_PRA_1983} using laser excitation from the X$^2\Sigma_g^+$ ground state.
The required excitation energy is given by the C$_2{}^-$ detachment threshold (3.269~eV \cite{Ervin_Lineberger_JPC_1991}).
Resonances in the electron energy spectra were mainly assigned to AD from levels in the B$^2\Sigma_u^+$ state (vibrational quantum number $v\geq5$) and indicate AD lifetimes of $\lesssim$\,$10^{-8}$\,s.
In fact, considering also radiative decay \cite{Rosmus_Werner_JCP_1984}, all B$^2\Sigma_u^+$ levels are expected to live for $<\SI{0.1}{\micro\second}$.
For AD from levels in the X$^2\Sigma_g^+$ and  A$^2\Pi_u$ states of C$_2{}^-$, high vibrational excitation ($v\geq16$ \cite{Ervin_Lineberger_JPC_1991}) is necessary and the processes were hardly observable so far.
As a source of the prominent 3\,ms AD signal observed on excited C$_2{}^-$, these levels are unlikely as their population would require extremely high vibrational temperature in view of the high detachment threshold.
The theoretically predicted \cite{Barsuhn_IOP_1974,Thulstrup_CPL_1974,Zeitz_CPL_1979,Shi_CTC_2016,da_silva_transition_2024} lowest quartet state of C$_2{}^-$, a$^4\Sigma_u^+$, thus logically enters as a further possible source of the stored-beam C$_2{}^-$ AD signal.
This was already speculatively considered \cite{Pedersen_JCP_1998,Iizawa_JPS_Japan_2022}.
However, to our best knowledge, no further specification about the role of this state in the AD of excited C$_2{}^-$ was given previously.
Moreover, we are not aware of any experimental spectroscopic characterization of the quartet state in free C$_2{}^-$.

Our theoretical study is partly based on earlier investigations on the stability of rotating anions \cite{Golser_ADH2_PRL_2005,Cizek_ADH2_PRA_2007,Marion_CU_PRA_2023}.
Similar to the methods of that work, our approach explicitly allows for electronically unstable initial anion states of resonant character, using a diabatic representation and employing the non-local resonance model \cite{Domcke_PR_1991,Cizek_HD_JPB_1998}.
This method fundamentally differs from that appropriate for the AD rates from doublet C$_2{}^-$ \cite{Mead_C2_JCP_1985}, based on non-adiabatic perturbation of adiabatic molecular levels as introduced by \citet{Berry_JCP_1966}.

In a companion paper (CP) \cite{PRA} we present experimental results on electron emission (AD) as well as heavy-particle autofragmentation (AF) using C$_2{}^-$ ions from a hot ion source injected into a cryogenic ion storage ring.
Moreover, we develop a detailed model of C$_2{}^-$ at high rotational and vibrational excitation that includes the radiative decay as well as the AD and AF rates for all relevant levels in the X$^2\Sigma_g^+$, A$^2\Pi_u$ and a$^4\Sigma_u^+$ states.
Together with a model of the level populations, these data complement the rates of rotationally assisted AD calculated here and allow us to show that AD and AF from a$^4\Sigma_u^+$ C$_2{}^-$ ions can plausibly explain the observed signals of both types in their intensities and their temporal behaviour.
Below, we shortly summarize the experimental AD signal and then present our theoretical method to assess the stability and the rotationally assisted AD decay of quartet  C$_2{}^-$.

\begin{figure}[t]
{\centering\includegraphics[width=0.50\textwidth]{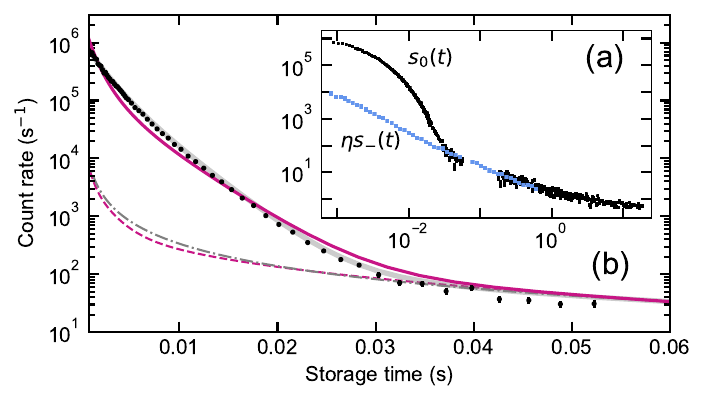}}
\caption{\label{fig-exper}
Count rates by electron emission (AD) and dissociation (AF) from C$_2{}^-$ as functions of the storage time $t$ \cite{PRA}. 
(a) Signals $s_0$ and $\eta s_-$ (see the text) with $\eta=0.75$. 
(b) Neutral signal $s_0$, assigned to AD and AF, with data shown as symbols.
Thick grey line: fit using a bi-exponential time dependence ($\tau_{1,2}$) of the AD signal and including a background (dashed, grey) from AF varying as $\propto t^c$ with $c=-1.287$, resulting from a fit to $\eta s_-$ at $t<0.06$\,s in (a). 
Purple lines: sum $s_d+\eta s_f$ of modeled AD and AF from C$_2{}^-$ in the a$^4\Sigma_u^+$ state (solid) and modeled AF background $\eta s_f$ (dashed-dotted).}
\end{figure}

The detected AD and AF count rates from a stored C$_2{}^-$ beam following its injection into a storage ring are shown in Fig.\ \ref{fig-exper}.
These data were obtained at the cryogenic storage ring CSR at the Max Planck Institute for Nuclear Physics in Heidelberg, Germany, as described in the CP \cite{PRA}.
The ions are produced in a sputter ion source and accelerated to 60~keV.
About $4\times10^{7}$ ions are then injected into the ring and stored.
Products of unimolecular decay are separated from the stored beam at the deflectors of the storage ring by their differing charge-to-mass ratios.
In separate runs, a single counting detector behind one of the deflectors was placed to intersect either neutral fragments (C$_2$ or C, count rate $s_0$) or C$^-$ fragments ($s_-$).
Products following electron emission (AD) contribute to $s_0$ only, while each dissociation event (AF) contributes to $s_0$ and $s_-$.
The cryogenic cooling of the storage ring leads to extreme vacuum such that C$_2{}^-$ destruction in collisions with residual gas molecules occurs too rarely to contribute significantly to the count rates.
Also, overall ion beam loss is expected to be insignificant during the displayed storage time interval.

The rate $s_0$ of neutrals shows a strong initial peak not appearing in $s_-$, thus corresponding to the AD signature.
The temporal dependence of this AD component is close to exponential, consistent with the time constant near 3\,ms observed earlier
\cite{Pedersen_JCP_1998,Iizawa_JPS_Japan_2022}.
We extract the AD component as $s_0-\eta s_-$ with a near-unity factor $\eta$ compensating intensity fluctuations between the runs \cite{PRA}.
This component is well fitted by a bi-exponential decay with time constants $\tau_1=3.31(4)$\,ms and  $\tau_2=1.56(5)$\,ms.

The complete picture of coexistent decay by AD, AF, and radiative relaxation is discussed in the CP \cite{PRA}.
It turns out that for levels showing AD, decay by AF is mostly negligible.
Moreover, radiative decay can significantly compete with AF, but is typically insignificant with respect to AD.
The decay constants of the experimental AD component, $\tau_{1,2}$, therefore reflect observed lifetimes against AD.

\begin{figure}[t]
\centering{\includegraphics[width=0.48\textwidth]{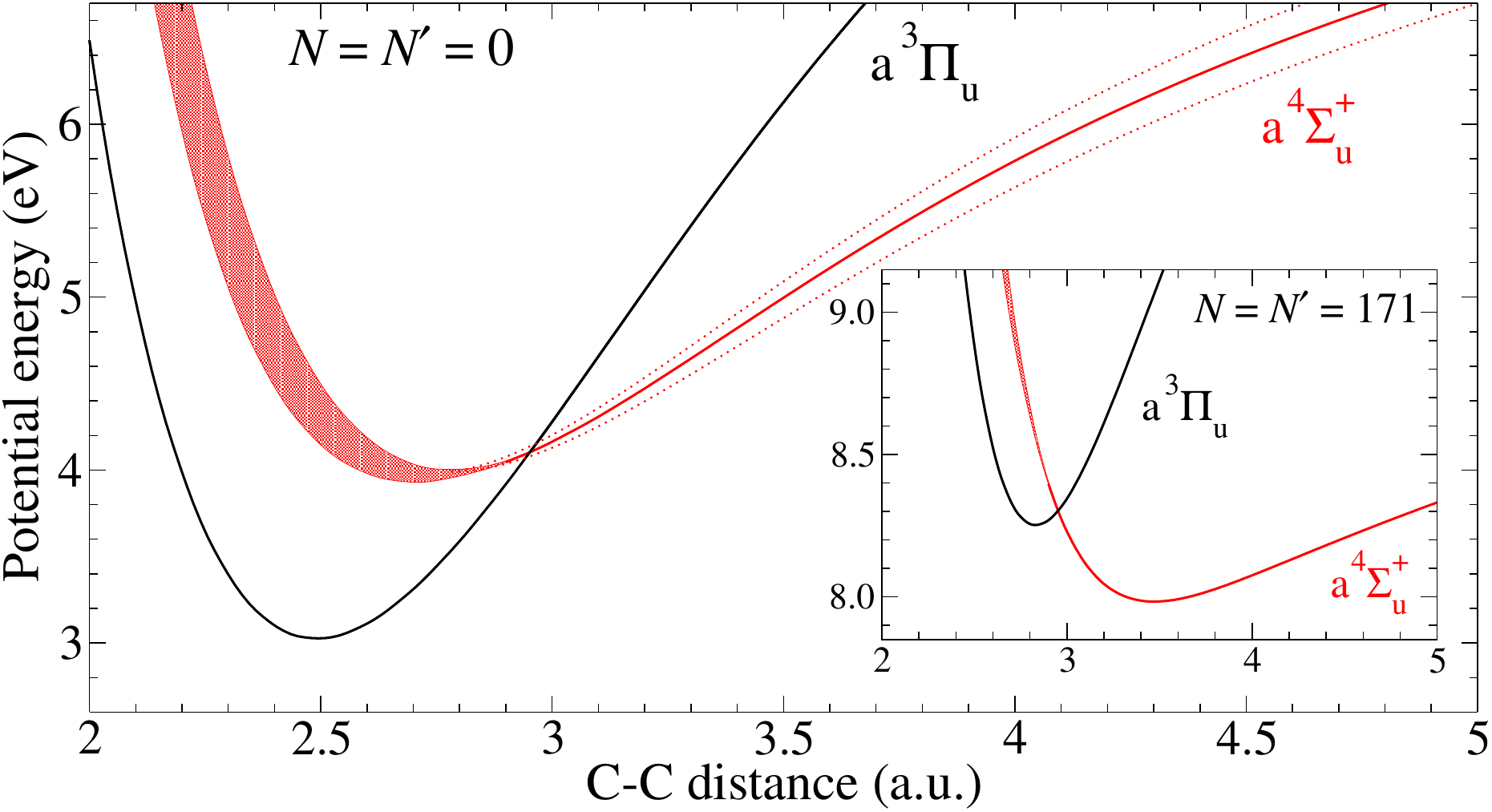}}
\caption{\label{fig-milan}
Potential curves of (red) C$_2{}^-$(a$^4\Sigma_u$) and (black) C$_2{}$(a$^3\Pi_u$) referred to the C$_2{}^-$ rovibrational ground state, neglecting nuclear rotation. Red-shaded: $R$-dependent width of the a$^4\Sigma_u$ AD resonance; dotted: \textit{ab initio} uncertainty of the bound-state potential.  Inset: Potential curves including the centrifugal term of Eq.\ (\ref{eq-Heff}) setting $N=171$ for the anion and $N'=N$ for the neutral.}
\end{figure}

Neglecting nuclear rotation, the C$_2{}^-$(a$^4\Sigma_u^+$) state can decay by spin-allowed AD into C$_2$(a$^3\Pi_u$).   
While predicted in multiple theoretical studies \cite{Barsuhn_IOP_1974,Thulstrup_CPL_1974,Zeitz_CPL_1979,Halmova_JPB_2006,Shi_CTC_2016,da_silva_transition_2024} this quartet anionic state could never be unambiguously assigned to any feature measured on free C$_2{}^-$ \cite{bondybey_photophysics_1975,weltner_carbon_1989,Pedersen_JCP_1998,Iizawa_JPS_Japan_2022}. 
For the present work, \textit{ab initio} calculations were performed \cite{PRA} using the {\sc molpro} package~\cite{MOLPRO12} to consistently locate this and the three lower lying anionic states together with the first two neutral states of C$_2$.  
Rotationless potential energy curves of the anionic a$^4\Sigma_u^+$ and neutral a$^3\Pi_u$ states are shown as a function of the internuclear distance $R$ in Fig.\ \ref{fig-milan}.    
For some $R$ range, C$_2{}^-$(a$^4\Sigma_u^+$) lies above C$_2$(a$^3\Pi_u$) and forms an electronic resonance. 
The resonance width at the molecular equilibrium position in the absence of nuclear rotation amounts to 0.48\,eV \cite{Halmova_JPB_2006} and in this case enables very fast AD with pico- to femtosecond lifetimes.
In the bound-state part ($R > 3$\,a.u.)\
we complemented the available resonance calculations \cite{Halmova_JPB_2006,Halmova_thesis} by our a$^4\Sigma_u^+$ \textit{ab initio} results whose uncertainty reaches $\pm 170$ meV at $R = 5$\,a.u.\ \cite{supp} (see Fig.~\ref{fig-milan}).
\nocite{Herzberg1_1950,OMalley_PR_1966,Chang_Fano_1972,Bieniek_1978,Haxton_H20_PRA_2007,Tarana_PRA_2009,Morgan_Chen_Rm_1997,Hazi_PRA_1979,Macek_PRA_1970,MOLPRO12,Dunning_pXZ_1989}

The arrangement of the curves changes for high rotations because of the significant centrifugal potential.
For a C$_2{}^-$ nuclear rotation quantum number $N\geq 155$ \cite{PRA} and assuming the quantum number $N'$ of the neutral molecule to be $N'=N$, the $v=0$ level of the anionic a$^4\Sigma_u^+$ curve lies below that of the neutral a$^3\Pi_u$ curve.
This form of a rearrangement of the neutral and anionic curves for higher rotational states was previously documented for $N=20$--40 of the H$_2{}^-$ and D$_2{}^-$ ions \cite{Golser_ADH2_PRL_2005,Marion_CU_PRA_2023,Jordon-Thaden_H2-_PRL_2011}.
The present, heavier molecule with a smaller rotational constant requires much higher $N$ for a similar rearrangement.
Considering, e.g., $N=N'=171$ (Fig.\ \ref{fig-milan}), the a$^4\Sigma_u^+$ ($v=0$) energy clearly lies below all $v$-levels of the neutral triplet state.
Quartet levels up to a certain $v$ then cannot undergo AD.
However, such rotationally stabilized levels are in fact not absolutely stable against AD, since processes with $N'<N$ for high enough $|N'-N|$ become energetically allowed again. 
This calls for the AD rate calculations to include such changes of the nuclear rotation.

A theory describing rotationally assisted AD of C$_2{}^-$ requires combining two approaches. 
The first one was used to predict the rotationally stabilized anion states of H$_2{}^-$ and its isotopologues \cite{Golser_ADH2_PRL_2005,Marion_CU_PRA_2023}. 
However, this technique did not allow for a change of the electronic angular momentum $l$ and kept it fixed in the $p$ wave. 
When applied to the present case, such limited rotational change of the molecular frame during the AD process predicts lifetimes $<${}$10^{-10}$\,s for the a$^4\Sigma_u^+$ levels.
A here incorporated second step corrects these shortcomings by accounting for the rotational degrees of freedom within the non-local resonance model \cite{Domcke_PR_1991}. 
Its derivation for $\Sigma$-states of anion and neutral can be found in Ref.~\cite{Cizek_Houfek_book_2012}.  
A similar procedure, based on the local complex potential was developed for the description of the angular distributions of the molecular fragments resulting from the dissociative electron attachment process
\cite{Haxton_h2o_h2s_PRA_2006}.

Along this line, we extend the non-local resonance model underlying the work by \v{C}\'{\i}\v{z}ek \textit{et al.}\ \cite{Golser_ADH2_PRL_2005,Cizek_ADH2_PRA_2007,Cizek_Houfek_book_2012,Marion_CU_PRA_2023}.
For the non-dissociative states $\ket{\Phi}$ and the narrow resonances, the diagonalization of the effective complex Hamiltonian $H^\mathrm{eff}$ provides the resonance energy $E$ as
\begin{equation}
\label{eq-H0}
H^\mathrm{eff}(E) \ket{\Phi} = E \ket{\Phi} \;.
\end{equation}
The energy dependence of $H^\mathrm{eff}$ leads to an iterative procedure that usually converged within a few iterations in the present study.
The effective nuclear Hamiltonian $H^\mathrm{eff}$ plays a central role in projection-operator methods that have been previously used to study AD \cite{Cizek_ADH2_PRA_2007,Golser_ADH2_PRL_2005}, dissociative electron attachment
\cite{Houfek_HCl_PRA_2002,Hotop_Advances_2003,Haxton_h2o_h2s_PRA_2006,Zawadzki_HNCO_PRL_2018}, vibrational excitation  \cite{Rescigno_CO2_PRA_2002,Cizek_HF_JPB_2003,Ragesh_HNCO_PRA_2020}, and associative detachment \cite{Miller_H2_PRA_2012,Roucka_PCL_2015}. 
In its local form it can be written, for each of the initial anion's rotational states $N$, as \cite{supp}
\begin{equation}
\label{eq-Heff}
H^\mathrm{eff}_N = T + \frac{N(N+1)}{2\mu R^2} + V_r(R) - \frac{i}{2}\Gamma_N(R)\;.
\end{equation}
Here $T$ is the radial nuclear kinetic energy operator and $V_r(R)$ the real part of the anion's potential.
Moreover,
\begin{eqnarray}
\nonumber
\Gamma_N(R) = \sum_{l N' \nu' \eta'} &\phantom{i}g^{\eta'}_{lN'N} (2N'+1) \left(
\begin{array}{ccc}
l & N' & N \\
\Lambda & -\Lambda & 0 \end{array} \right)^2  \\
\label{eq-GammaR}
&\times \Gamma_l(R)\, |\chi_{\nu'N'}(R)|^2 \;,
\label{eq:gamman}
\end{eqnarray}
is the $R$-dependent total width, where $\Gamma_l(R)$ is the local partial width for partial wave $l$, $\chi_{\nu'N'}(R)$ describes the final rovibrational states of the neutral molecule, and the factor $g^{\eta'}_{lN'N}$ = 0 or 1 selects only one of the two final, nearly degenerate a$^3\Pi_u$ states (hence, $\Lambda$ = 1) in order to preserve the nuclear spin state during the AD process \cite{supp}. 
The imaginary part, Eq.\ (\ref{eq-GammaR}), of $H^\mathrm{eff}$, Eq.\ (\ref{eq-Heff}), clearly shows how the lifetime of the initial anion state (a$^4\Sigma_u^+,N$) is determined by the final rovibrational state $(v' N')$ and by the rate, represented by $\Gamma_l(R)$ \cite{supp}, at which the electron is ejected in the partial wave $l$.

\begin{figure}[t]
\begin{center}
\includegraphics[width=0.48\textwidth]{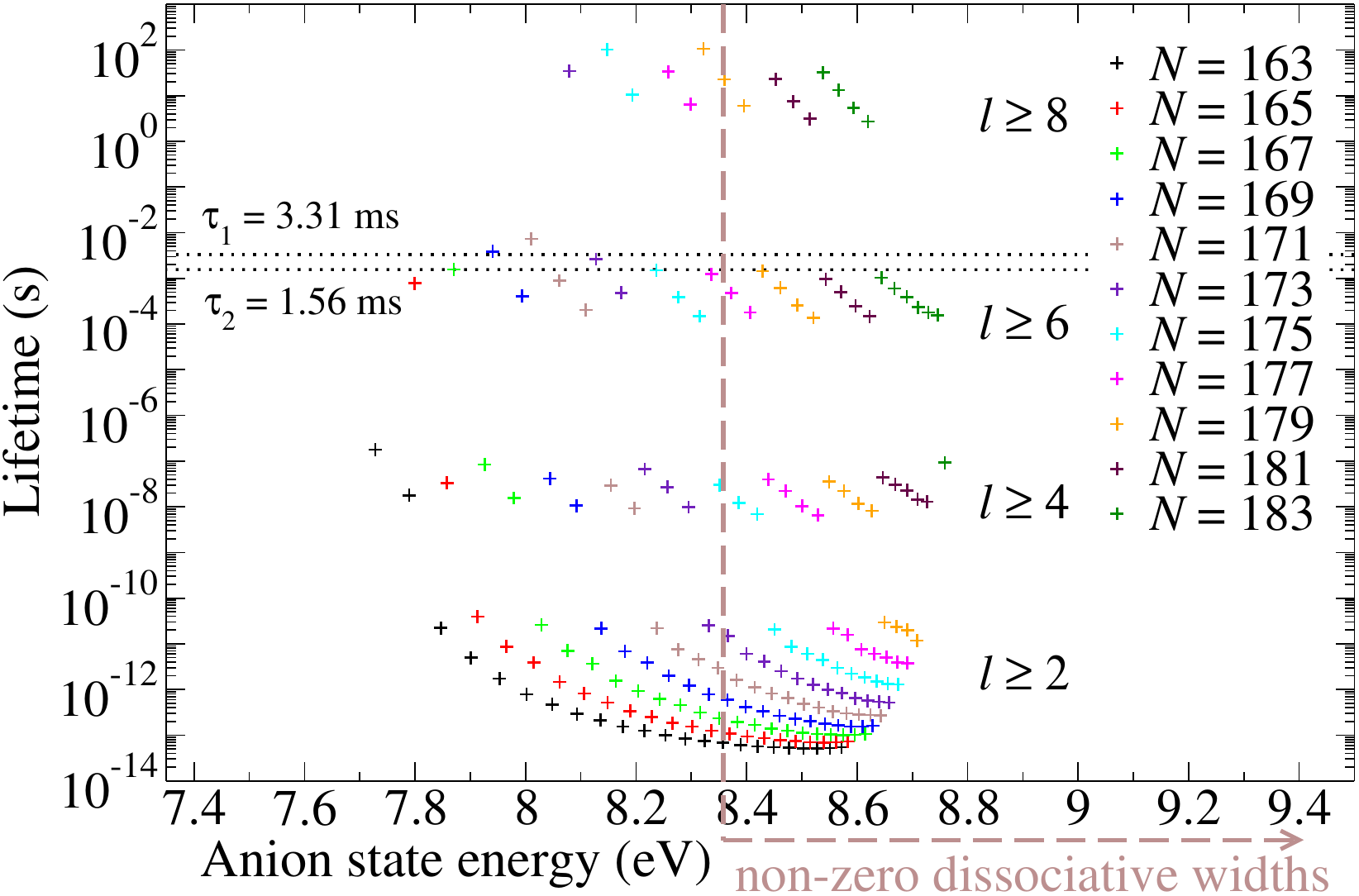}
\caption{\label{fig-lifet}
Energies and lifetimes $\tau_{\text{AD}}$ of anion states $v,N$, color-coded according to $N$, with the partial wave ($l$) range of separated groups, the threshold for tunneling predissociation (long-dashed line), and the experimental time constants $\tau_{1,2}$.
}
\end{center}
\end{figure}

The initial and final states are both treated as Hund's case (b), since for the a$^3\Pi_u$ state both the spin--orbit coupling constant \cite{Ervin_Lineberger_JPC_1991} and its rotational dependence \cite{Curtis_Sarre_JMS_1985} are small enough that for the considered strong rotation, spin-orbit energy shifts can be neglected.
The splitting of the final a$^3\Pi_u$ states due to the $\Lambda$-doubling is $<3$ meV up to $N'=180$ \cite{Curtis_Sarre_JMS_1985,PRA} and neglected in the present study.

The quartet level energies and vibrational wave functions used in averaging $\Gamma_N(R)$ of Eq.\ (\ref{eq:gamman}) were in a first step obtained with the centroid (see Fig.\ \ref{fig-milan}) of the \textit{ab initio} a$^4\Sigma_u^+$ potential curve \cite{supp}.
Calculated lifetimes for various initial $N$-levels are shown in Fig.~\ref{fig-lifet}.
The lifetimes of levels within a narrow $N$-range spread over many orders of magnitude, as was similarly seen for the H$_2{}^-$ calculations \cite{Golser_ADH2_PRL_2005,Cizek_ADH2_PRA_2007}.
Unlike these previous results, there are also well-separated groups with sets of lifetimes similar to each other on the logarithmic scale.
Closer inspection reveals that the levels in such groups share the lowest angular momentum $l$ of the ejected electron. 
\begin{figure}[tb]
\begin{center}
\includegraphics[width=0.48\textwidth]{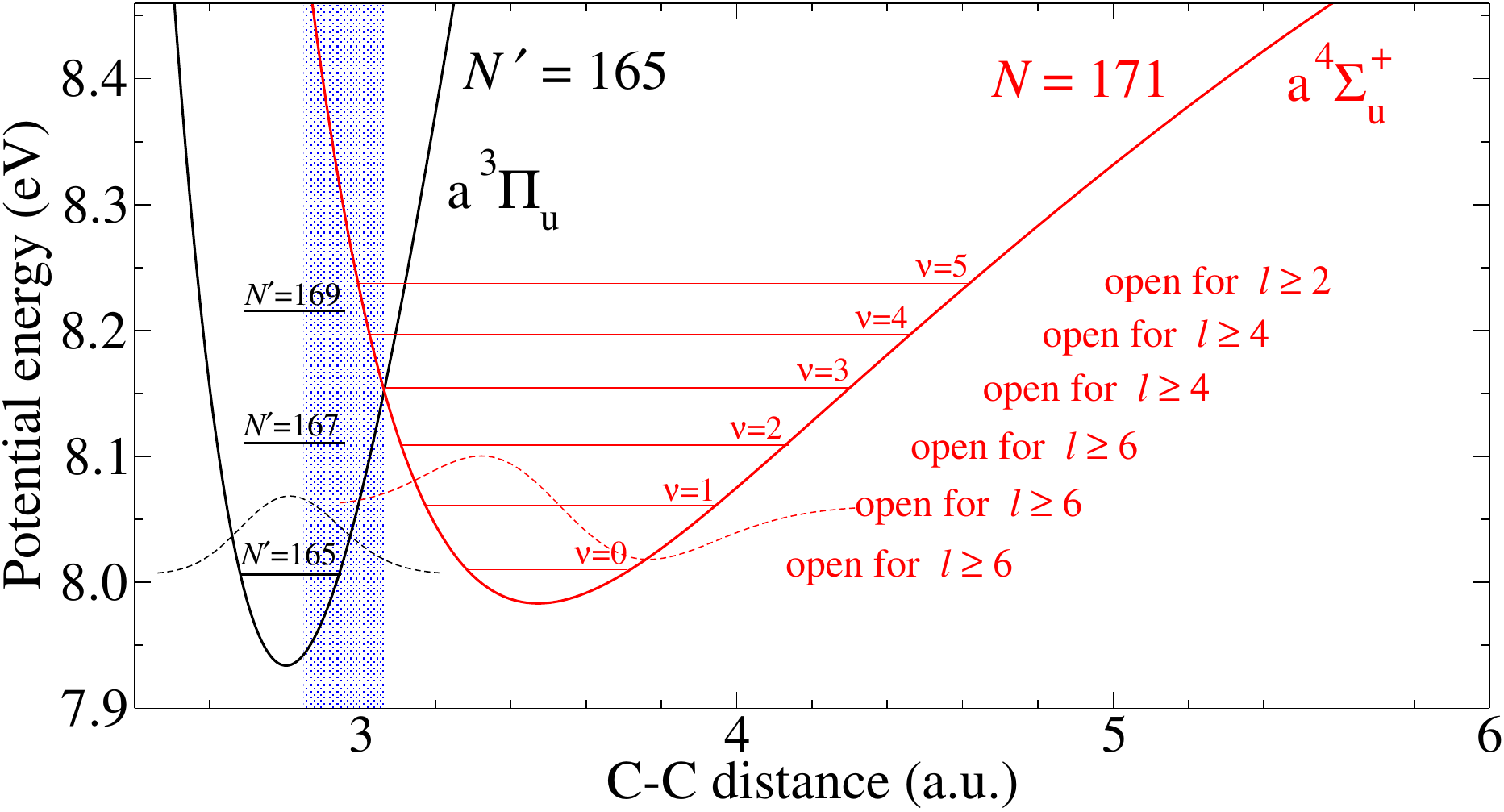}
\caption{\label{fig-pots_171}
Potential curves and sample vibrational wave functions (dotted) for AD with $N=171$ and $N'=165$ and relevant zone of vibrational overlap (shaded).  Also shown are anionic $v$-levels with the $l$-range for which AD is energetically allowed and some $v=0$ neutral levels as $N'$ increases.}
\end{center}
\end{figure}
We emphasize the following aspects of the results:

(i) Lifetimes of the experimental order of magnitude ($\sim$\,10$^{-3}$\,s) are 
found for states that require $l\geq6$ for AD.
Such states are only available for a narrow interval of initial rotational states $N=165$--183.
For lower $N$, rearrangement is not strong enough to forbid the fast AD through $l=2,4$ partial waves (see Fig.\ 1 of the SM \cite{supp}), while  higher $N$ do not host suitable bound vibrational levels.

(ii) The $N'$ values required to open AD for the $v$-levels of given $N$ are illustrated for $N=171$ in Fig.\ \ref{fig-pots_171}.
The levels $v\leq2$ require $N'<167$ 
and hence $l\geq6$ from Eq.\ (\ref{eq:gamman}).
In the indicated vibrational overlap zone, $\Gamma_N$ is thus controlled by $\Gamma_6(E)$.
Its small value reflects the threshold law $\Gamma_l(E) \propto E^{l+1/2}$ \cite{Domcke_PR_1991}, resulting in millisecond AD lifetimes for the $N=171, v=0,1,2$ levels.

We use these calculated lifetimes $\tau_{\text{AD}}(v,N)$ against AD for the partial wave $l=6$ to model the experimental neutral count rate $s_0=s_d+\eta s_f$ from highly excited C$_2{}^-$.
The time constants $\tau_{\text{AD}}(v,N)$ 
are implemented in the C$_2{}^-$ unimolecular decay model of the CP \cite{PRA}, yielding $s_d(t)$ for AD and $s_f(t)$ for AF.
Deriving rotational and vibrational level populations, we use bi-modal thermal distributions compatible with available evidence on sputter ion sources \cite{PRA}.
For the included distribution tails, this evidence suggests temperatures of $\lesssim${}$\SI{1e4}{\kelvin}$.
When we respect this limit within the model \cite{PRA}, the a$^4\Sigma^+_u$ levels turn out to be the only ones able to yield the observed AD signals. 
The set of $\tau_{\text{AD}}(v,N)$ for $l=6$ 
and a parameter $P_q$ giving the population fraction of a$^4\Sigma^+_u$ C$_2{}^-$ anions with AD lifetimes of $>$\,$10^{-6}$\,s
determine the temporal trend.

The \textit{ab initio} uncertainty of the bound a$^4\Sigma^+_u$ potential acts on the $\tau_{\text{AD}}(v,N)$ values. 
Closest agreement with the experimental signal, as shown in  Fig.\ \ref{fig-exper}(b), is reached when this potential is increased to the upper margin of the uncertainty range, and for $P_q =6\times10^{-3}$.
When other regions within the uncertainty range are chosen, the AD decay model becomes dominated by levels with somewhat longer lifetimes \cite{PRA}.
The model also shows that, for the same thermal distribution, the AF signal ($s_f$) stems from the tunneling dissociation of rotationally stabilized C$_2{}^-$ quartet levels at high $v$ (but at lower vibrational energy and, thus, lower vibrational temperature than would be required for C$_2{}^-$ doublet levels).
Correspondingly, the modeled signals $\eta s_f$ and $s_d+\eta s_f$ are both shown in Fig.\ \ref{fig-exper}(b), where $s_d$ mostly dominates over $\eta s_f$. 
Since AF strongly depends on vibrational excitation, other, less vibrating quartet levels with negligible AF and very long AD lifetimes in the range for $l\geq8$ (Fig.\ \ref{fig-lifet}) should exist.

Using the model, we have also probed the excitation conditions of the  C$_2{}^-$  beam which could explain the observed AD and AF signals without a quartet C$_2{}^-$ population \cite{PRA}.
Then, tail temperatures of $\SI{3e4}{\kelvin}$ (carrying $>$\,3.5\% of the population) are required.
We consider such a scenario highly unlikely and see strong evidence for the presence of quartet levels in the investigated  C$_2{}^-$ ion beams.

In summary, we present a theory that predicts the AD of a diatomic molecule when it is connected with changes of rotation by as many as six quanta.
The correspondingly long AD lifetimes become relevant for rotationally stabilized quartet levels of the C$_2{}^-$ anion and reproduce the puzzling AD signals with long decay times observed on stored ensembles of this anion.
The derived dependence of the AD width on the number of rotational quanta exchanged during the process is expected to be of general use for future treatments of strongly rotating molecular anions. 

\begin{acknowledgments}
This article comprises parts of the doctoral thesis of V.C.S. submitted to the Ruprecht-Karls-Universit\"at Heidelberg, Germany. The work of R.\v{C}. has been supported by the Czech Science Foundation (Grant No.\ GACR 21-12598S). Financial support by the Max Planck Society is acknowledged. The computational results presented have been in part achieved using the HPC infrastructure LEO of the University of Innsbruck.
\end{acknowledgments}

\newpage\null\thispagestyle{empty}\newpage

\includepdf[pages=1]{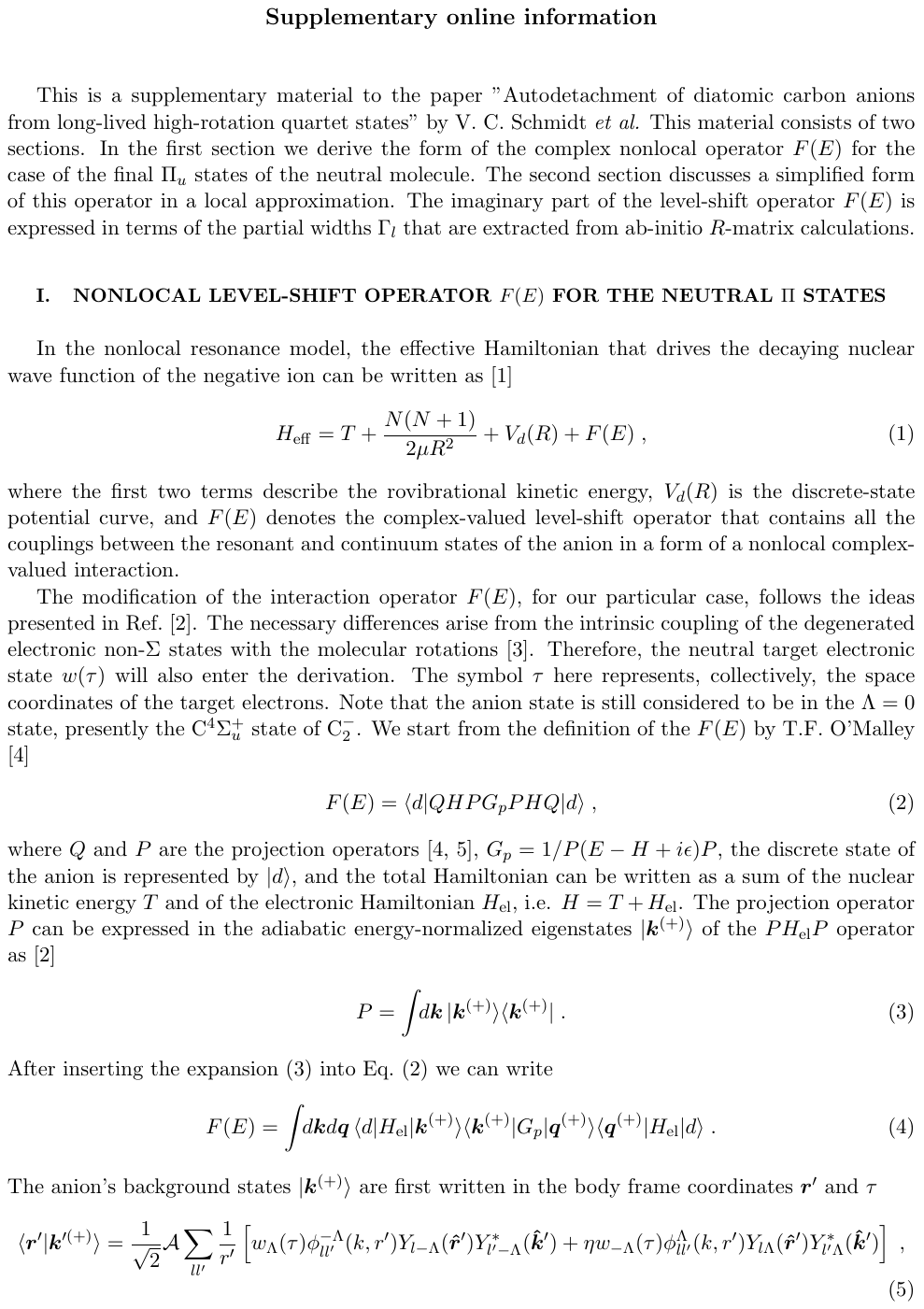}\null\thispagestyle{empty}\newpage
\includepdf[pages=2]{suppl.pdf}\null\thispagestyle{empty}\newpage
\includepdf[pages=3]{suppl.pdf}\null\thispagestyle{empty}\newpage
\includepdf[pages=4]{suppl.pdf}\null\thispagestyle{empty}\newpage
\includepdf[pages=5]{suppl.pdf}\null\thispagestyle{empty}\newpage
\includepdf[pages=6]{suppl.pdf}\null\thispagestyle{empty}\newpage
\includepdf[pages=7]{suppl.pdf}

\end{document}